\documentclass[11pt]{article}
\usepackage{hyperref}

\usepackage{graphicx}
\usepackage{epsfig}
\usepackage[numbers]{natbib}
\usepackage{textcomp}
\usepackage{amssymb}

\begin{document}
\title{{\bf Fundamental Physics with Charged Particle Measurements at the Cherenkov Telescope Array}}
\author{J.~Vandenbroucke, B.~Humensky, J.~Holder, R.~A.~Ong}
\date{\today}

\maketitle

\section{Charged cosmic particles and fundamental physics}

Cosmic electrons, positrons, protons, and antiprotons carry essential information about both astrophysical processes and fundamental physics.  Charged particles provide signatures of dark matter annihilation or decay, \emph{in situ} in the universe, that are complementary to the other messengers of indirect dark matter detection, gamma rays and neutrinos.  We discuss the science potential for measurements of charged particles by the Cherenkov Telescope Array (CTA), an instrument under development primarily for the study of cosmic gamma rays with energies $30\ \textrm{GeV} - 200\ \textrm{TeV}$.

The cosmic-ray antiproton spectrum has been measured from balloons and satellites between 60 MeV and 180 GeV~\cite{PhysRevLett.105.121101}.  Secondary antiprotons are created as cosmic rays propagate through the galaxy via several processes, such as $pp \rightarrow p\bar{p}pp$, and the flux of these secondary antiprotons is very sensitive to the propagation processes of cosmic rays. Above ~10 GeV, the ratio of antiprotons to protons from secondary production is expected to decline. Various new-physics scenarios can lead to an increase in the antiproton to proton ratio at energies above 100 GeV, with predictions up to $\sim10^{-2}$ for contributions from dark matter~\cite{Cirelli20091} and up to several percent for contributions from extragalactic sources such as anti-galaxies~\cite{1984Natur.309...37S}.  Antiprotons can also be produced by primordial black hole evaporation~\cite{refId0}.  Existing measurements up to $\sim 180\ \textrm{GeV}$ are consistent with diffuse secondary production.  At higher energies, ground-based experiments featuring instrumentation to detect air shower particles at ground level have set upper limits on antiprotons using the shadow of the Moon~\cite{PhysRevD.85.022002}.  The Cherenkov Telescope Array will have sufficient effective area to make an important contribution between 200 GeV and a few TeV. 

Electrons and positrons can likewise be produced by secondary production processes during propagation of cosmic-ray nuclei through the Galaxy, or in the vicinity of astrophysical sources such as supernova remnants and pulsars.  They can also be among the final-state products from annihilation or decay of dark matter particles. The combined electron plus positron spectrum has been measured between 7~GeV and 1~TeV by the Fermi satellite~\cite{2009PhRvL.102r1101A}, and up to 6 TeV by HESS~\cite{2008PhRvL.101z1104A} and MAGIC~\cite{2011ICRC....6...43B}.  Moreover, in the past few years a surprising discovery (confirming earlier hints from balloon-based experiments) has been made in cosmic positrons: the PAMELA, Fermi, and AMS satellites~\cite{2009Natur.458..607A,PhysRevLett.108.011103,PhysRevLett.110.141102} determined that the positron fraction increases with energy between 10~GeV and 350~GeV, while the prediction from astrophysical secondary production is that the fraction should decrease with energy.  Many models have been developed to explain this surprising excess of positrons, the most popular of which invoke either pulsars or dark matter.  Extending these measurements to as high energy as possible is essential to discriminating among these models.

\section{Future measurements with CTA}


Over the next decade, several balloon- and satellite-based instruments will make important measurements of antiprotons, electrons, and positrons.
Fermi and CREST will measure the total electron plus positron spectrum up to several TeV~\cite{2008ICRC....2..305S}, and
CALET will measure it to $\sim$20~TeV~\cite{2013AIPC.1516..293R}.  
AMS will extend its positron fraction measurement beyond 350~GeV, precisely measure the total electron plus positron spectrum to $\sim$1~TeV or beyond, and provide a precise measurement of antiprotons.


CTA will complement these satellite- and balloon- based instruments with its large effective area at very high energies.  With its excellent energy and direction reconstruction capabilities as well as good ability to distinguish hadronic and electromagnetic showers, CTA will make a precise measurement of the total electron plus positron spectrum between $\sim$100~GeV and $\sim$100~TeV.  A large fraction of the sky (directions without bright gamma-ray sources) can be used.  CTA can also provide strong constraints on large-scale anisotropies (an important signature in distinguishing source models) in the electron plus positron spectrum by comparing flux measurements made in opposite directions on the sky~\cite{2013arXiv1304.1791L}.  Furthermore, CTA can use the method of the Moon shadow combined with the geomagnetic field to act as a spectrometer, to measure or constrain both the antiproton and positron spectra in the region between several hundred GeV and $\sim$2~TeV.  This technique was pioneered by the ARTEMIS collaboration using the Whipple 10~m telescope~\cite{2001APh....14..287P}.

Depending somewhat on the geomagnetic field strength at the CTA site\footnote{CTA site selection is underway.}, the lunar shadow as detected in singly charged particles is displaced from the lunar direction by $\sim$1.5\textdegree~at 1 TeV, corresponding to a total separation of $\sim$3\textdegree (compared with the Moon diameter of $\sim$0.5\textdegree) between particles of charge +1 and particles of charge -1.  The magnitude of this deflection is equal for electrons, positrons, protons, and antiprotons in this ultra-relativistic regime.  

While the Moon spectrometer technique is promising, there are several experimental challenges.  These are being addressed with data taking campaigns by the current generation of imaging atmospheric Cherenkov telescopes (IACTs).  The most significant challenge is to understand the operation and performance of Cherenkov telescopes pointed close to the Moon during phases with significant moonlight.  Among the current generation of IACTs, MAGIC began a program to search for the electron and positron shadows during the 2010-11 season, and anticipates it will take several years to acquire the estimated 50 hours of observations needed to see the electron shadow~\cite{2009arXiv0907.1026C,2011ICRC....6..189C}. VERITAS has developed UV-passing filters that can be placed in front of the cameras to suppress a significant fraction of the background light from the Moon and has begun a multi-year observing campaign during the 2012-13 season.

CTA, with an order of magnitude better sensitivity than the current generation of instruments, is expected to detect the electron shadow in the 300 to 700 GeV range in under 5 hrs and  will have correspondingly improved sensitivity for the positron and antiproton measurements.  Depending on the positron flux in this range, it will likely take several times larger integration time to detect the positron shadow than the electron shadow.

The US contribution to CTA is particularly important for these scientific goals.  The US proposal is to contribute an array of mid-size telescopes with 8\textdegree~diameter field of view and cameras instrumented with silicon photomultipliers (SiPMs).  Compared to the CTA design without the US contribution, the US contribution of mid-size telescopes provides three important improvements for these measurements.  First, the US contribution will increase the point-source sensitivity in the core energy range (100~GeV to 10~TeV) by a factor of two to three~\cite{2012AIPC.1505..765J}.  The energy range covered by these mid-size telescopes is the range where CTA can contribute best to the positron spectrum measurement.  At lower energies the satellite-based experiments have better sensitivity, and at higher energies the positron and electron shadows are not as well separated and the fluxes are significantly lower.  Second, SiPMs are not endangered by bright moonlight (as PMTs are), so there are less stringent restrictions on pointing the telescopes near or at the Moon.  Observations can therefore be made under a wider range of conditions (during greater moon phase), enabling a greater integrated sensitivity to be achieved.  Third, the improved optical point spread function and smaller pixel size of the innovative dual-mirror design planned for the US contribution enables improved hadronic/electromagnetic shower discrimination.  This is important because hadronic showers are much more numerous than electron and positron showers and even a small fraction of residual background can present a significant challenge to these measurements.



Instruments on balloons, satellites, and on the ground will provide important new measurements of charged cosmic particles over the next decade, extending the surprising discoveries that have generated a great deal of excitement among particle physicists in the past few years.  These new measurements are essential to distinguishing among the many competing models.  Charged particles are an important component of indirect dark matter searches, complementing gamma-ray and neutrino measurements.  CTA will contribute unique measurements of cosmic electrons, positrons, and antiprotons at the highest energies.

\bibliography{snowmass_charged_particles}		

\end{document}